# Waiting for Help: Timely Access to Psychological Support for Young Adults Exposed to Parental Substance Misuse

Bastien Michel, Søren Albeck Nielsen, Morten Hesse,

Kristine Rømer Thomsen, and Marianne Simonsen*

March 2026

Access to mental health care is often rationed through waiting lists, yet there is limited causal evidence on the consequences of delayed access. We study whether eliminating waiting time for psychological support improves outcomes for young adults who grew up with parental substance misuse. Using a randomized waitlist-controlled trial in Denmark combined with survey and administrative data, we find that immediate access leads to sizable short-run improvements in psychological health. These gains persist three to four years after randomization, even after both groups have received the intervention. By contrast, we find limited evidence of large average effects on broader health or labor market outcomes. Our results highlight the importance of treatment timing in capacity-constrained settings.

* Bastien Michel (bastien.michel@univ.nantes) is affiliated with Nantes University (LEMNA). Søren Albeck Nielsen (snielsen@econ.au.dk) and Marianne Simonsen (msimonsen@econ.au.dk) are affiliated with the Department of Economics and Business Economics at Aarhus University. Morten Hesse (mortenhesse@gmail.com) and Kristine Rømer Thomsen (krt.crf@psy.au.dk) are affiliated with the Department of Psychology and Behavioural Sciences at Aarhus University. We would like to thank TUBA (and, in particular, Henrik Appel, Alex Kastrup Nielsen, Marc Krabbe-Sørensen, Helle Lindgaard, and Henrik Prien) for their support to the research project since the start of the project. We thank Maj Hansen for sharing the validated Danish version of the International Trauma Questionnaire prior to publication and to Sanne Dalgaard Toft for outstanding project management. Finally, we would also like to thank Helena Skyt Nielsen, Lars Skipper, and seminar participants at Aarhus University and Nantes University, as well as participants at FEW 2025 and AHEPE 2025 workshops and the AFEPOP 2024, ESPE 2024, and SOLE 2025 conferences. Financial support from Det Obelske FamilieFond, Lundbeck Fonden, Novo Nordisk Fonden, and TrygFonden is gratefully acknowledged. Michel received additional financial support for this project from the Pays de la Loire region through the PULSAR program. The study received IRB approval from Aarhus University (serial number 2019-17) and was registered in the AEA RCT Registry (AEARCTR-0005776). The findings, interpretations, and conclusions expressed in this paper, as well as all errors, are entirely ours.
Corresponding author: bastien.michel@univ-nantes.fr

# I. Introduction

Access to mental health care is often rationed through waiting lists. This is especially true for specialized services targeted at vulnerable populations, where demand frequently exceeds capacity (OECD 2020). As mental health needs rise across high-income countries, waiting times have become a central policy concern. Yet there is limited causal evidence on the consequences of delayed access to care—particularly for individuals with substantial prior exposure to adversity. Understanding these effects is critical for policy design, as waiting time may itself constitute a central and understudied margin of health care rationing.

In this paper, we study whether timely access to psychological support improves outcomes for young adults who grew up with parental alcohol or drug misuse. We focus on individuals aged 25–35, a group that is largely overlooked in existing interventions, which typically target children and adolescents still living in affected households. Specialized support for this population is scarce, and there is limited evidence on how best to assist them.

Globally, an estimated 400 million people suffer from alcohol use disorders (WHO 2024) and around 64 million from drug use disorders (UNODC 2025). While the health and social consequences of these behaviors are severe for the individuals involved—substantially increasing the risk of violence, self-harm, disease, and mortality (e.g., Griswold et al. 2018)—a growing literature documents important spillover effects on others. In particular, recent work on the opioid epidemic shows that substance misuse has far-reaching consequences for children's environments and long-run outcomes (Evans, Lieber, and Power 2022; Buckles et al. 2023).

A large literature shows that children of substance-misusing parents face elevated risks of adverse psychological and behavioral outcomes (Johnson and Leff 1999; Christensen and Bilenberg 2000; Lindgaard 2002, 2005; Smith et al. 2016). They are also more likely to experience disability, engage in self-destructive behavior, and face an increased risk of premature mortality (Christoffersen and Soothill 2003). One potential mechanism is that parental substance misuse increases the likelihood of involuntary out-of-home placements, which may themselves have important implications for later outcomes (Helénsdotter 2026).

Despite this evidence, much less is known about whether—and when—interventions can mitigate these risks once individuals reach adulthood. This is a consequential gap. Early adulthood is a critical period for human capital accumulation and family formation: individuals are entering the labor market, forming long-term relationships, and becoming parents themselves. At this stage, unresolved psychological distress may have persistent consequences not only for individual well-being, but also for labor market attachment and intergenerational outcomes.

We study a large Danish provider of free psychological counseling and therapy for individuals affected by parental substance misuse. Because capacity is constrained, access to services is governed by waiting lists that in some locations exceed one year. We exploit this setting through a randomized waitlist design. Young adults who sought help in Denmark's four largest cities and faced waiting times above one year were randomly assigned either to begin therapy immediately or to receive access after a one-year delay. We combine this experiment with survey data and population-wide administrative registers and follow participants for up to four years after randomization.

Our design identifies the effect of timely access to care rather than the effect of treatment receipt. Individuals assigned to delayed access were free to seek other forms of support while waiting, and many did so. The comparison is therefore between immediate access to the focal intervention and the set of outside options realistically available during a long wait. This makes the setting directly relevant for policy environments in which mental health care is capacity constrained and waiting time is a key margin of rationing.

The paper delivers three main findings. First, take-up of immediate access is very high, indicating strong engagement with the intervention among help-seeking individuals. Second, eliminating the one-year wait produces sizable short-run improvements in self-reported psychological health. One year after randomization, individuals offered immediate access report higher mental well-being and lower levels of psychological distress, depression, and post-traumatic stress symptoms. Third, some of these gains persist in the medium run. Two to three years after the delayed-access group became eligible for treatment, the group initially assigned to immediate access still exhibits better psychological outcomes. By contrast, we find limited evidence of large average effects on broader health or labor market outcomes.

This paper makes three contributions. First, it provides causal evidence on whether timely access to psychological support can mitigate the long-run consequences of adverse childhood experiences in adulthood, complementing a literature that has largely focused on childhood exposures (e.g., Currie et al. 2022; Cabral et al. 2026; Helénsdotter 2026). Second, it contributes to the literature on health care rationing and waiting times (e.g., Hamilton and Bramley-Harker 1999; Cullis et al. 2000; Koopmanschap et al. 2005; Nikolova et al. 2016; Reichert and Jacobs 2018; Williams and Bretteville-Jensen 2022; Godøy et al. 2024; Costantini 2025) by showing that delays in access can generate persistent losses even when treatment is eventually received. Third, it contributes methodologically by combining a randomized waitlist design with population-wide administrative data to study policy-relevant constraints in a real-world setting with minimal attrition. More broadly, the results speak to ongoing concerns about long waiting times for mental health services in OECD countries (OECD 2020).

The remainder of the paper proceeds as follows. Section II describes the institutional setting and the intervention. Section III presents the experimental design and conceptual framework. Section IV outlines the empirical strategy. Section V describes the data. Section VI presents the results. Section VII concludes.

## II. Support for young adults from substance-affected families

II.A Context

In Denmark, alcohol misuse is a widespread problem and, consequently, many individuals are growing up or have grown up in families concerned by this issue. In total, it is estimated that 9.5 percent of children are growing up in families with alcohol problems (Kristiansen et al. 2008). Although the prevalence of drug misuse is lower, it generates comparable adverse consequences within families.

II.B The institutional setting and generalized support

To grasp the consequences of early provision of help from TUBA, it is important to understand the institutional context, including access to the usual care or support systems typically available in the absence of TUBA.

In Denmark, individuals can always seek help via their primary care provider, whose services are fully funded by the public health insurance. Primary care providers engage in primary disease prevention and health maintenance as well as diagnosis and treatment of minor acute and chronic illnesses. Crucially, they additionally serve as gatekeepers to private practicing specialists such as psychologists and psychiatrists at a reduced cost (usually, a referral gives access to a 60 percent subsidy but may entail considerable wait times of sometimes up to a year[1]), just as they can refer patients to secondary hospital-based psychiatric care in more severe instances. This can be provided as both outpatient and inpatient care and is free of charge but again often associated with some wait period.[2]

Yet, the traditional care system may not be perfectly adapted to the needs of adult children of substance misusing parents, who may have difficulties trusting and accepting care from others (Straussner and Fewell, 2011) and, often refrain from seeking help from traditional healthcare providers. Additionally, the available support—such as that offered by private-practice psychologists—is often not tailored to the specific needs of this population, whose distress initially stems from their close relationship with someone suffering from substance misuse or a substance use disorder, rather than from a clinical diagnosis of their own. However, if left unaddressed, such distress can escalate into more serious mental health conditions, underscoring the need for early, targeted support.

II.C Targeted support from TUBA

TUBA was established in 1997 in Denmark with the objective of helping children and young adults from substance-affected families. It operates through 33 centers geographically distributed across the country and involves around 181 licensed therapists (approximately 30 percent salaried employees and 70 percent volunteer workers). TUBA offers different types of interventions (e.g., individual therapy, group therapy, and online therapy). Participation is entirely voluntary and at the initiative of those in need of help.

TUBA's main intervention consists of tailored mainstream individual psychotherapy, free of charge for the clients.[3] TUBA follows an approach that integrates the principles of so-called humanistic and psychodynamic therapy. In general terms, their therapists help individuals focus on their personal concerns, values, interests, and goals while also recognizing the significance of challenges experienced in their childhood home and how these may continue to affect them today.

---

[1] https://www.sundhed.dk/borger/guides/find-behandler/, accessed March 18, 2026.
[2] https://www.esundhed.dk/Emner/Patienter-og-sygehuse/MitSygehusvalg, accessed March 18, 2026.
[3] Mainstream psychotherapy has been shown to effectively reduce psychological problems across a wide range of treatment approaches and conditions (Barkham et al., 2021).

Upon registration, help seekers undergo a screening process. Eligible individuals are placed on a waiting list,[4] while those deemed ineligible (e.g., individuals experiencing psychosis or dealing with substance misuse/substance use disorders) are redirected to more appropriate assistance. While the Danish Authority of Social Services and Housing provides funding as part of a broader initiative to ensure short waiting times at TUBA for children from substance-affected families under the age of 25,[5] this support is not available to older individuals leading to waiting times of over a year in the country's four largest cities (Copenhagen, Odense, Aalborg, and Aarhus).[6]

Once the waiting period is over, clients are assigned a therapist who follows them throughout the intervention. Sessions are held in person at one of TUBA's centers, with the frequency of appointments mutually determined by the therapist and client.

### III. The experimental design and conceptual framework

III.A The experimental design

In March 2019, we launched a randomized controlled trial designed to measure the impact of eliminating waiting time for TUBA's program. We focus on young adults aged 25 to 35 who sought support in large cities and, as previously mentioned, faced extended waiting time of more than one year. Data collection ran for over four years and ended in August 2023. Appendix Figure A1 summarizes the project timeline, detailing the launch of the trial, the rollout of treatment, and the completion of follow-up data collection.

*Sampling:* From March 2019 to March 2020, we drew a sample of 358 respondents from individuals who had sought help at one of TUBA's four largest centers (Copenhagen, Odense, Aalborg, and Aarhus), where the waiting time for individuals registering exceeded one year throughout the study period.

Eligibility for study inclusion was assessed based on information collected during the routine screening meeting that help-seekers have with one of TUBA's therapists upon registration. We only considered individuals for study inclusion who were deemed eligible to TUBA's help. In line with TUBA's usual practice, people in circumstances deemed incompatible with the conduct of the intervention (e.g., those with own drinking problems) were referred to more appropriate services and not considered for study inclusion. Additionally, individuals who required urgent support were provided with immediate assistance (until they were fit to join the waiting list) and not considered for study inclusion either.

---

[4] TUBA receives funding from municipal grants as well as via donations from various foundations. Still the organization is unable to provide help to everyone in our target group immediately and an increasing number of individuals are placed on waiting lists.
[5] https://www.sbst.dk/projekter-og-initiativer/boern/gratis-behandlingstilbud-til-boern-og-unge-fra-familier-med-stof-eller-alkoholmisbrug, accessed March 18, 2026.
[6] In 2022, an additional 559 individuals joined the waiting lists in Denmark's four largest cities—Copenhagen, Odense, Aalborg, and Aarhus—amid steadily increasing delays. For example, between 2021 and 2023, waiting times rose from 16 to 24 months in Aarhus and from 25 to 30 months in Copenhagen.

Respondents were recruited in four batches over a 13-month period. The first batch was made up of eligible individuals who were on the waiting list in March 2019 and faced a waiting time exceeding one year. The three other batches were made up of eligible help seekers who enrolled between May 2019 and March 2020. All individuals contacted for study inclusion were informed that they would have a 50 percent chance of starting treatment immediately and a 50 percent chance of starting in a year's time. Participation in the study was entirely voluntary, and respondents were informed that they could withdraw at any time. In total, the first batch included 150 individuals, the next two 70 individuals each, and the last one 68 individuals.

*Randomization:* In each study site, we randomly selected half of the respondents of each batch to begin the intervention immediately (the treatment group), while the other half was scheduled to start the treatment a year later (the control group). This ensured that, at randomization, individuals who were offered to start the intervention immediately were comparable to those who were placed on the waiting list and would be offered to start the intervention a year later. TUBA had no involvement in the randomization process.

*Ethical considerations:* The study received IRB approval from Aarhus University. Two important ethical considerations deserve to be emphasized in the design of the experiment. First, since participation in the study was offered only to individuals who, at the time of registration, were facing a waiting period of over a year, participation in the experiment did not extend the waiting time for control group respondents. Second, the project was made possible through generous donations from several foundations, enabling TUBA to recruit additional therapists. This allowed more individuals to access the intervention simultaneously, without increasing waiting times or compromising the quality of therapy, including for those not participating in the study.

*Other considerations:* It should be noted that throughout the project, the implementation of the study was conducted by all therapists at the four study sites, thus ensuring that the study results reflect the work of TUBA as a whole, and not just that of a few therapists. Finally, note that during the height of the COVID pandemic, some therapy sessions were held online.[7]

III.B Conceptual framework

Given the nature of the intervention, we expect that early access primarily affects help-seekers' psychological health. TUBA's services are tailored to individuals exposed to parental substance misuse and may therefore be more effective than alternative forms of support available during the waiting period. This is particularly relevant in a setting where all participants have actively sought help, suggesting a high degree of engagement with the intervention.

We expect the effects of early access to be larger in the short run—before the control group receives treatment—than in the longer run. However, effects may persist even after delayed individuals gain access for at least two reasons. First, delays may lead to a deterioration in psychological health that subsequent treatment does not fully reverse (a level effect). Second, waiting may reduce motivation or beliefs about the possibility of improvement, lowering the returns to treatment once it is received (a production function effect). At the same time, if improvements in psychological health fade over time, treatment effects may attenuate and could converge once the control group begins therapy.

Finally, improvements in psychological health may translate into broader outcomes. To the extent that the intervention enhances individuals' ability to cope with everyday challenges, we may observe changes in health behaviors and the use of other services. The direction of effects on health care utilization is theoretically ambiguous: improved well-being may reduce the need for care, while increased awareness and agency may lead to higher take-up. We therefore view effects on these outcomes, as well as on labor market attachment, as empirical questions.

III.C Data

Our study relies on rich data from multiple sources, including administrative records from TUBA, participant survey responses,[8] and register-based data from Statistics Denmark.

Surveys were conducted to gather information on respondents' psychological health and behaviors throughout the study period. Baseline survey data were collected as part of TUBA's screening procedure prior to randomization. Two follow-up surveys were then administered. The first follow-up was conducted one year after treatment respondents were offered to start the intervention, and just before control respondents were given the option to begin. The second follow-up survey was carried out between May and August 2023, three to four years after treatment respondents were offered the opportunity to start the intervention and two to three years after control respondents were offered access to help.[9] Survey data were collected through online questionnaires.

---
[7] A limited lockdown was imposed in March 2020, during which all public sector employees in non-essential roles were ordered to stay home for two weeks.
[8] See again Appendix Figure A1 for information about the dates of the various surveys for the four batches of respondents.

We supplement self-reported survey data with two types of administrative data. First, we use administrative records from TUBA, linked through respondents' client ID, to document the intervention's implementation. These records provide information on the dates of all sessions attended by help seekers, enabling us to identify treatment start dates, duration, and the total number of sessions completed. Second, we utilize register-based data from Statistics Denmark, linked via respondents' Central Personal Registration (CPR) numbers—a unique identifier assigned to all Danish residents, which we collected at baseline. This allows us to obtain additional baseline and follow-up information, particularly regarding respondents' health and labor market attachment. We were able to successfully match survey data with register-based administrative records for 98 percent of the individuals in the sample.

**Main outcomes**

Our pre-registered primary outcome is respondents' psychological health, measured using four validated instruments. Together, these measures capture complementary dimensions of psychological health, including well-being, distress, depression, and trauma-related symptoms.

First, we rely on the World Health Organization-Five Well-Being Index (WHO-5) to measure mental well-being (WHO 1998, 2024). The index captures how frequently individuals experienced positive feelings over the past two weeks, such as feeling "happy and in a good mood" or "active and vigorous," and ranges from 0 to 100, with higher values indicating greater well-being.

Second, we draw on the Clinical Outcomes in Routine Evaluation-Outcome Measure (CORE-10) to assess psychological distress (Barkham et al. 2013). This measure captures the frequency of negative experiences over the past week, including feelings of being "overwhelmed by problems" or "terribly alone and isolated," and ranges from 0 to 40, with higher values indicating more severe distress.

---

[9] By design, the time between the start of the intervention for respondents in the treatment and control groups and the second follow-up survey varies from one batch to another.

Third, we employ the Major Depression Inventory (MDI) to measure symptoms of depression over the past two weeks (Bech et al. 2001). The index reflects the frequency of symptoms such as "feeling low in spirit or sad" or "decreased appetite," and ranges from 0 to 50, with higher values indicating more severe depressive symptoms.

Fourth, we use the International Trauma Questionnaire (ITQ) to measure symptoms of post-traumatic stress disorder (PTSD) (Hyland et al. 2017; Cloitre et al. 2018; Hansen et al. 2021). Respondents report the extent to which they are affected by issues related to a traumatic event, such as "upsetting dreams" or "avoiding reminders," with scores ranging from 0 to 24 and higher values indicating more severe symptoms.

For each measure, we also construct indicators based on established clinical thresholds to capture the prevalence of severe psychological distress.

**Additional outcomes**

We also examine broader outcomes related to health and socio-economic behavior. To capture health-related behaviors, we use survey data from the CAGE-C questionnaire to measure alcohol-related problems (Hjarnaa et al. 2023), including questions on drinking frequency and whether respondents feel a need to reduce their alcohol consumption.

We further study the use of health care services using register-based data on psychotropic drug consumption and contacts with health care providers. Psychotropic drug use is measured in Defined Daily Doses (DDD), the World Health Organization standard unit capturing average daily consumption.

Finally, we examine labor market outcomes using register-based data. These include employment status (any employment and full-time employment), earnings, and the receipt of public benefits.[10]

---

[10] We also pre-registered charges and criminal convictions as possible secondary outcomes. However, these outcomes turned out to be extremely rare among sampled individuals.

## IV.    Estimation strategy

IV.A Main analysis

We measure the impact of the intervention on individuals using an Intention-To-Treat (ITT) approach. To do so, we estimate the following equation:

$$y_i = \beta T_i + \mu_i + X_i \gamma + \varepsilon_i \quad \textbf{Equation (1)}$$

where $y_i$ is the outcome measured at follow-up 1 or 2, $T_i$ is a dummy variable taking the value 1 if respondent *i* was offered to start the intervention immediately (treatment respondents) and 0 if they were placed on a waiting list (control respondents), $\mu_i$ is a vector of strata fixed effects used for stratification (Bruhn and McKenzie, 2009), and $X_i$ is a vector containing baseline covariates. To discipline covariate selection and reduce researcher discretion in a high-dimensional setting, we employ the double-lasso procedure of Belloni et al. (2014).

In Equation (1), the coefficient of interest is β, which we estimate at two distinct points in time. First, we estimate β one year after randomization, just before the control group was granted access to the intervention. This estimate captures the short-term impact of being offered immediate access to treatment relative to remaining on a waiting list for one year. Second, we estimate β at the time of the 2nd follow-up survey, three to four years post-randomization. It captures the longer-term effect of the intervention, two to three years after control group participants were granted access to the intervention.

For all estimates of β, we report two-sided p-values from a randomization inference test under the null of zero treatment effects. This procedure reassigns treatment status using the original stratified randomization procedure and re-estimates β under 2,000 placebo assignments. The p-value corresponds to the share of placebo estimates with an absolute value greater than or equal to the observed estimate. This procedure allows for robust inference, especially in cases of small sample sizes.

IV.B Heterogeneity analysis

To better understand our main results and their implications, we also examine the impact of the intervention across different pairs of mutually exclusive subgroups by estimating the following equation:

$$y_i = \beta_1(Group_{1,i} * T_i) + \beta_2(Group_{2,i} * T_i) + \gamma Group_{1,i} + \mu_i + X_i\beta + \varepsilon_i \quad \textbf{Equation (2)}$$

In this specification, $Group_{1,i}$ and $Group_{2,i}$ are indicator variables denoting whether respondent $i$ belongs to group 1 or group 2 (e.g., female or male respondents). Again, $X_i$ represents a vector of baseline covariates selected via a double-lasso procedure. These heterogeneity analyses were not pre-registered and should be interpreted cautiously.

We examine heterogeneous treatment effects by sex, baseline psychological health, and intention to have children to explore whether immediate access may help reduce the intergenerational transmission of unresolved trauma. For psychological health, we split the sample into halves based on baseline measures of mental well-being and PTSD symptom intensity.

To shed light on potential intergenerational implications, we examine heterogeneity by parenthood status and life stage. We first compare individuals with and without children at baseline. While this distinction captures realized fertility rather than intentions, it provides a natural proxy for exposure to parenting responsibilities and decisions. Given the relatively small share of parents in the sample (24 percent of 358 participants), we complement this analysis by comparing older (aged 30 and above) and younger respondents. Age captures proximity to family formation and may therefore proxy for differences in fertility intentions, although it may also reflect broader differences in maturity and life circumstances. We interpret these heterogeneity analyses as suggestive evidence on whether timely access to support matters differentially for individuals at stages of life where intergenerational transmission is most relevant.

## V. Descriptive statistics and validity tests

V.A Descriptive statistics: characterizing help-seekers

We start by exploring the characteristics of the study participants to learn about their vulnerabilities and strengths. Table 1, Columns 1-2 summarizes information on respondents at baseline. In line with behaviors observed across health care more generally (Mustard et al., 1998; Arendt, 2012), the vast majority (74 percent) of the respondents were women. Individuals were on average 30 years old at baseline and were all raised in a household where an adult had an alcohol misuse problem; 37 percent also reported that an adult had a drug problem. On average, respondents reported becoming aware of the problem at the age of 11, 80 percent of respondents said they had experienced violence in childhood, and 37 percent had a family member who had attempted suicide. At baseline, 62 percent reported poor mental well-being, 84 percent had symptoms of mild to severe psychological distress, 51 percent reported symptoms of mild to severe depression, and 38 percent experienced PTSD. In total, 14 percent had attempted suicide prior to seeking help. At the time of randomization, 64 percent were employed, 24 percent were enrolled in education, and 12 percent were neither in education nor employment.

For a perspective on external validity, Appendix Table A1 shows the profile of individuals seeking help from TUBA, relative to the general population in the same age group. Help-seekers exhibit weaker attachment to the labor market, with earnings that are 29 percent lower on average. Additionally, they rely more heavily on public benefits and make greater use of the primary healthcare system. They are also 15 percentage points less likely to be married and 19 percentage points less likely to have children.

Since many of our outcomes are based on self-report data, a potentially critical issue for our formal analysis is survey attrition, particularly for the second follow-up survey, conducted three to four years after randomization. Fortunately, the completion rate for the first follow-up survey is high, at 94 percent, and remains substantial at 72 percent for the second follow-up survey. Table 1, Columns 3-6 present the characteristics of respondents who completed the first follow-up survey and those who responded to the second. A comparison of these groups' characteristics with the baseline respondents also suggests that attrition did not result in significant changes to the composition of the sample (Columns 7 to 9): differences in baseline characteristics across samples are small and rarely statistically significant.

V.B Balance checks

Table 2 continues to compare the average observable baseline characteristics of individuals who were offered immediate access to the intervention with those who were placed on a waiting list and offered the intervention one year later. To test differences across groups, we regress each baseline characteristic (listed in the leftmost column of the table) on a dummy variable indicating assignment to the treatment group, controlling for the vector of stratification fixed effects. Parameter estimates (*p*-values) from these regressions are shown in Column 2 (Column 3). Differences between groups are generally small and rarely statistically significant, suggesting that the randomization procedure successfully created comparable groups: of the 29 tests performed, two are statistically significant at the 10 percent level, one at the 5 percent level, and none at the 1 percent level.

Moreover, we find no evidence of differential attrition across groups at either of the two follow-ups, with neither difference being statistically significant at the 10 percent level. Additionally, we repeat the balance checks while restricting the sample to individuals who completed the first follow-up survey (Table 2, Columns 4–6) and the second follow-up survey (Table 2, Columns 7–9). Again, differences remain small and rarely statistically significant, confirming that attrition did not compromise the comparability of the treatment and control groups.

## VI. Results

VI.A Participation in the intervention and behaviors of individuals while wait-listed

We begin by examining respondents' help-seeking behavior throughout the study period, with a particular focus on their engagement with TUBA's intervention. We pay special attention to the behavior of control group participants while they were on the waiting list. To document these patterns, we draw on both TUBA's administrative records and survey data.

First, and maybe not surprisingly since participants actively sought help in the first place, there is high demand for the specialized psychological support offered by TUBA among the target population, confirming that the intervention addresses a previously unmet. Using TUBA's administrative data, Figure 1, Panel A, shows that take-up of the intervention rose rapidly among treatment respondents shortly after they were invited to start, such that, 12 months after randomization, all treatment participants had initiated the intervention (see Table 3, Panel A).

Second, while on the waiting list, a substantial share of control respondents sought support outside of TUBA. Twelve months after randomization, shortly before control participants were offered access to the intervention, 50 percent reported having received some form of assistance in the past year—47 percent from sources outside TUBA and 33 percent from formal external providers (Table 3, Panel

A). The most common formal sources of support were psychologists (22 percent) and psychotherapists (7 percent) unaffiliated with the organization (Figure 1, Panel B). In contrast, these figures are markedly lower among respondents randomized to immediate treatment: 28 percent reported receiving any additional support from outside TUBA, 9 percent consulted a psychologist, and only 1.1 percent saw a psychotherapist.

From a wellbeing perspective, control respondents who sought formal help during the 12-month waiting period differed significantly from those who did not. As shown in Appendix Table A2, those who sought formal help exhibited poorer baseline psychological health across all four validated instruments and had earnings 33 percent lower than their peers. This suggests that waiting lists may be particularly burdensome for individuals facing the most acute disadvantages, prompting them to seek formal support elsewhere despite the potential costs involved.

After waitlist-respondents were offered access to TUBA's intervention (12 months after randomization), participation rates rose sharply: according to TUBA's administrative data, 89 percent of them had initiated the intervention by the time of the second follow-up survey, three to four years after randomization (Table 3, Panel B). This highlights that the need for help remains exceptionally high, even among individuals who faced a substantial waiting period and further underscores the persistent nature of the challenges faced by our target population. Nevertheless, control respondents were 12 percentage points less likely to start the intervention once eligible (significant at the 5 percent level). These findings suggest that waiting time may reduce eventual uptake and increase the likelihood of receiving no support—mechanisms that we incorporate into our longer-term impact analysis.[11]

Finally, we find no evidence of differences in the intensity of participation between treatment and control respondents. In the control group, the average duration of participation was 18 months, during which individuals attended an average of 19 one-hour therapy sessions (Table 3, Panel B). The differences with treatment participants are not statistically significant at the 10 percent level.[12] This points to longer-term differences between groups not being due to differences in the intensity of treatment. From a policy perspective, this also indicates that the intervention was not particularly intensive—on average, one meeting per month.

---

[11] Note that Panel B reveals a difference between the full sample and the subsample that responded to the second follow-up survey, since in the latter all control individuals had received the intervention. This suggests that the longer-term effects measured using these data may represent only a lower bound of the treatment effect.
[12] It should be noted that TUBA does not issue any guidelines regarding the duration or number of sessions per client that would explain this result.

VI.B The short-term impact of immediate access to help

We measure the short-term impact of eliminating waiting time for TUBA's intervention on respondents' psychological health using survey data collected 12 months after randomization, just before control respondents were offered to start the intervention. We also use administrative data to measure any wider impact on respondents' lives. It is important to stress that these estimates do not reflect a comparison between receiving TUBA's intervention and receiving no support at all. Instead, as explained above, they capture the effect of the intervention relative to the counterfactual scenario in which individuals spend a year on the waiting list and during which half sought and obtained alternative forms of assistance. It should also be noted that this estimate does not reflect the effect of the intervention upon completion either, as its average duration (19 months) exceeds the 12-month follow-up period.

Table 4, Panel A shows that compared to being waitlisted for a year, being offered immediate access to the intervention substantially improved respondents' mental well-being as measured by the WHO-5 score by as much as 0.6 standard deviations. Early access to help also reduced symptoms of psychological distress, depression, and post-traumatic stress disorder by 0.33 to 0.39 standard deviations. All four coefficients are statistically significant at the one percent level. These conclusions are insensitive to whether or not baseline covariates are included in the estimated equation. In the control group, mental well-being remains largely stable during the year in which participants await support, while symptoms of psychological distress, depression, and PTSD decline slightly. This suggests minor improvements, potentially driven by control individuals seeking alternative forms of support while on the waitlist, as discussed further below.

Interestingly, as displayed in Figure 2 and Table A3, Panel A, the effects on the standardized scores of our psychological health measures are largely driven by reductions in the share of individuals classified as experiencing particularly poor psychological health. Among treatment respondents, the proportion reporting poor mental well-being declines from 69 to 46 percent (a 33 percent reduction); those experiencing at least moderate psychological distress fall from 50 to 27 percent (also a 33 percent decrease); and the share meeting the criteria for PTSD drops from 36 to 23 percent (a 36 percent reduction). For symptoms of depression, however, the intervention's primary effect lies in reducing the proportion of individuals with mild symptoms, with no statistically significant change among those with more severe forms. These improvements are primarily explained by increases in the proportion of individuals classified as having no psychological health problems (for mental well-being, depression, and PTSD) or only mild ones (for psychological distress).

However, as shown in Table 5, the improvements in respondents' psychological health did not translate into statistically significant effects across many other dimensions of their lives overall. Contrary to prior findings concerned with access to mental health treatment in a more general population (Prudon, forthcoming), we find no significant impacts on labor market attachment, including employment status 12 months after randomization, earnings, or the number of months during which respondents received any type of benefit. Note, of course, that in our population, many (24 percent) were enrolled in education and half received some type of income assistance at baseline; see Table 1. Treatment respondents were, however, eight percentage points less likely to report alcohol-related problems, as measured by the CAGE-C (statistically significant at the 10 percent level). These findings caution against assuming that gains in psychological well-being will necessarily produce large spillover effects in the short-run on economic or behavioral outcomes commonly targeted by policy.

VI.C The longer-term effects

We measure the longer-term impact of eliminating waiting time for TUBA's intervention using survey data collected two to three years after control respondents were offered to start the intervention, as well as register-based data.

Table 4, Panel B, indicates that immediate access to psychological support leads to lasting improvements in help-seekers' psychological health outcomes. Indeed, while control respondents' mental well-being had improved and symptoms of psychological distress, depression, and PTSD decreased when they were granted access to help, their psychological health does not catch up with that of respondents who were offered to start the intervention immediately. Two to three years after being offered to start the intervention, control respondents continued to exhibit worse psychological health outcomes: their overall mental well-being was 0.22 standard deviations lower (p-value=0.09), and their symptoms of depression and post-traumatic stress disorder were higher by 0.28 and 0.19 standard deviations, respectively (p-values=0.01 and 0.08).

Again, the improvements observed for the treatment group are largely accounted for by a significant reduction in the share of individuals in particularly poor psychological health; see Figure 2 and Table A3, Panel B. In particular, control respondents were eight percentage points more likely to experience moderate to severe psychological distress, eight percentage points more likely to experience symptoms of mild to severe depression, and 15 percentage points more likely to be diagnosed as experiencing symptoms of post-traumatic stress disorder – all three coefficients being statistically significant at the one percent level. Notably, the proportion of control group respondents remains largely unchanged between the first and second follow-ups, suggesting that delayed access to treatment may undermine the recovery potential of individuals with severe psychological health challenges, in contrast to immediate access to support. We also observe a concomitant increase in the proportion of people classified as having no or only mild psychological health problems, suggesting that in the longer term, eliminating waiting time for psychological support helps reduce the likelihood of people resolving their problems.

Again, we find no overall longer-term effects of eliminating waiting times for psychological support on labor market attachment or health care take-up—once control group respondents are eventually offered and, in most cases, have received the intervention.

Finally, note that these longer-term effects do not appear to be driven by sample selection related to the relatively higher attrition rate at follow-up. As previously discussed, attrition did not significantly alter the composition of the sample (Table 1) or introduce systematic differences between treatment and control groups (Table 2). In addition, we replicated the analyses presented in Tables 3 and 4 using the subsample of respondents who participated in both follow-up surveys and obtained similar results; see Appendix Tables A4a and A4b.

VI.D Heterogeneity analysis

We examine heterogeneous treatment effects by sex, baseline psychological health, and intention to have children. Results are reported in Tables 6 and 7, which present impacts at the time of the first and second follow-ups, respectively.

*By sex*: We find no statistically significant differences in psychological outcomes between male and female respondents in either the short or long run. However, we detect a modest positive effect on male's likelihood of full-time employment at the $2^{nd}$ follow-up, with no corresponding effect for females. This difference may reflect gender norms, with males having traditionally stronger attachment to the labor market—labor force participation rates in 2024 were 67.8 percent among males and 59.8 percent among females—and their greater propensity to recover from labor market shocks (Ivandić and Lassen, 2023).

*By baseline psychological health*: In the short run, early access to psychological support improves psychological well-being for both groups, with effect sizes of similar magnitude among those with better and poorer initial mental health. This indicates that, in the short run, the intervention benefits all, not only those with the lowest psychological health. In the longer run, although differences are not statistically significant, point estimates are considerably larger and more statistically significant among participants with the lowest baseline psychological health, suggesting that early intervention is particularly effective for those starting from a more vulnerable position. We find no significant impacts on labor market or broader health-related outcomes in either the short or long term, regardless of baseline psychological health.

*By intention to have children:* In the short run, improvements in psychological health are positive for both baseline parents and non-parents, as well as for younger (24–29) and older (30–35) respondents, with larger effect sizes for parents and older participants. In the longer run, these gains remain stronger and more statistically significant among these groups, although differences in ITT estimates across subgroups are not always significant. Although point estimates are somewhat more frequently statistically significant for older respondents, their magnitudes are similar among individuals who had children at baseline. This suggests that differences between parents and non-parents might have been more often statistically significant with a larger sample of parents. Overall, these results indicate that early psychological support may help reduce the intergenerational transmission of unresolved trauma. However, as in previous analyses, we find limited effects on labor market and other health outcomes in both the short- and longer-term.

## VII. Conclusions

This paper studies the causal effects of eliminating waiting times for psychological support among young adults who grew up with parental substance misuse. We show that timely access to specialized support leads to meaningful improvements in psychological health, both in the short run and over a longer horizon.

Individuals offered immediate access experience lower levels of psychological distress, depression, and post-traumatic stress symptoms, as well as a reduced prevalence of severe mental health problems. These effects arise in a setting where the intervention is targeted and relatively low-cost, highlighting the potential importance of access to appropriate support rather than treatment intensity per se.

By contrast, delayed access is associated with worse outcomes. Even after all participants have been offered treatment, individuals initially assigned to immediate access continue to exhibit better psychological health, suggesting that delays can generate losses that are not fully offset when care is postponed.

Taken together, our findings point to waiting time as an important margin of health care rationing. In capacity-constrained systems, reducing delays in access to psychological support may yield benefits that extend beyond the period of treatment itself. These considerations are particularly relevant given persistent waiting times for mental health services across OECD countries.

# Figures

**Figure 1:** Participation in treatment and behavior of individuals on waiting lists timeline

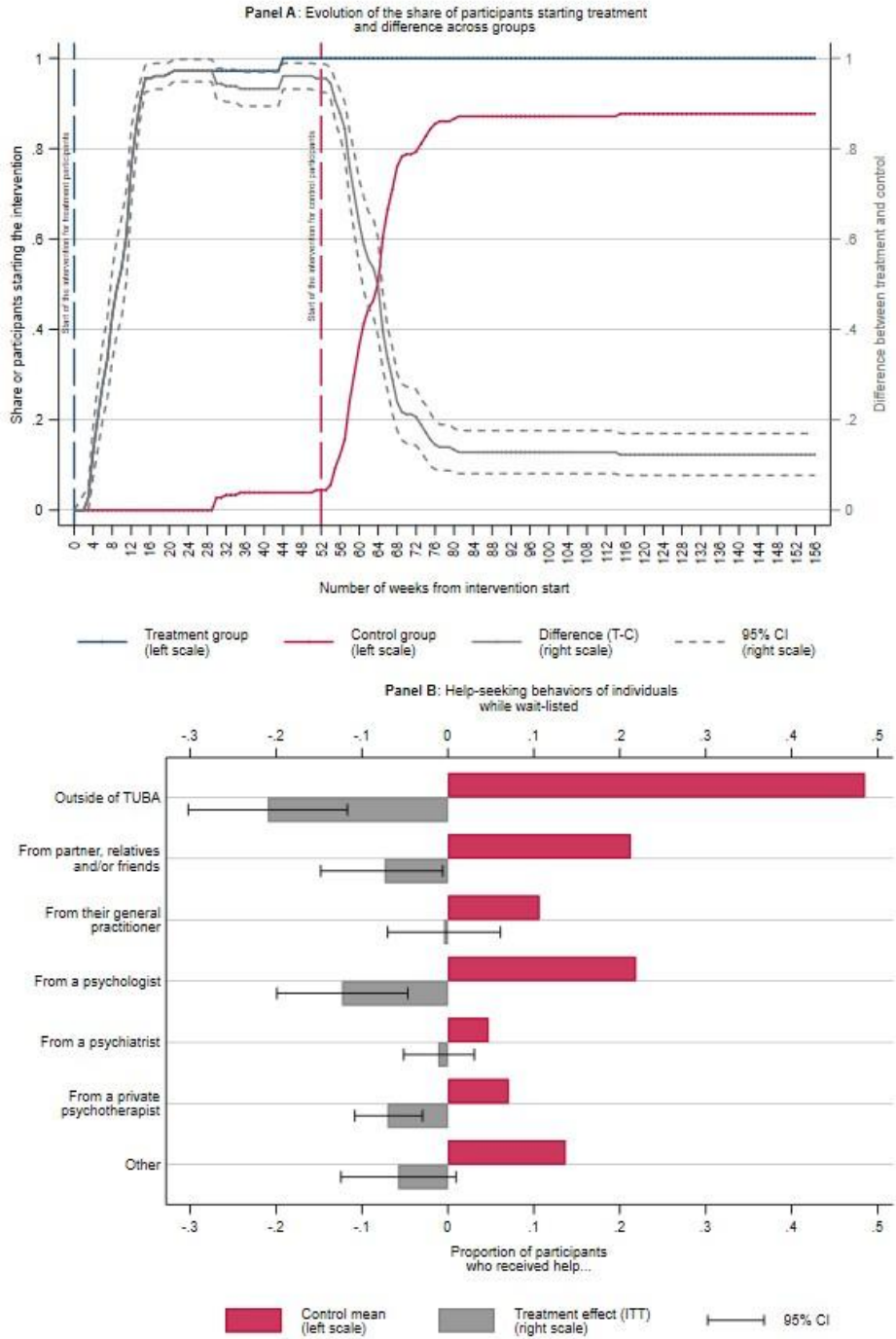

*Notes*: *Panel a)* shows the proportion of control and treatment respondents who had started the intervention at the end of each week from randomization until three years later (left-hand scale). We also show the difference between the proportion of respondents from the treatment and control groups who started the intervention at the end of each week over the same period (right scale) (N=358). *Panel b)* shows the share of respondents who declared having received help in the form of counselling or therapy from anyone or any organization *other than* TUBA, as well as the nature of the help received (N=337). Figures are provided separately for control and treatment respondents. Confidence intervals are obtained by wild bootstrap (2,000 replications).

**Figure 2:** Impact of the intervention on respondents' psychological health (categories)

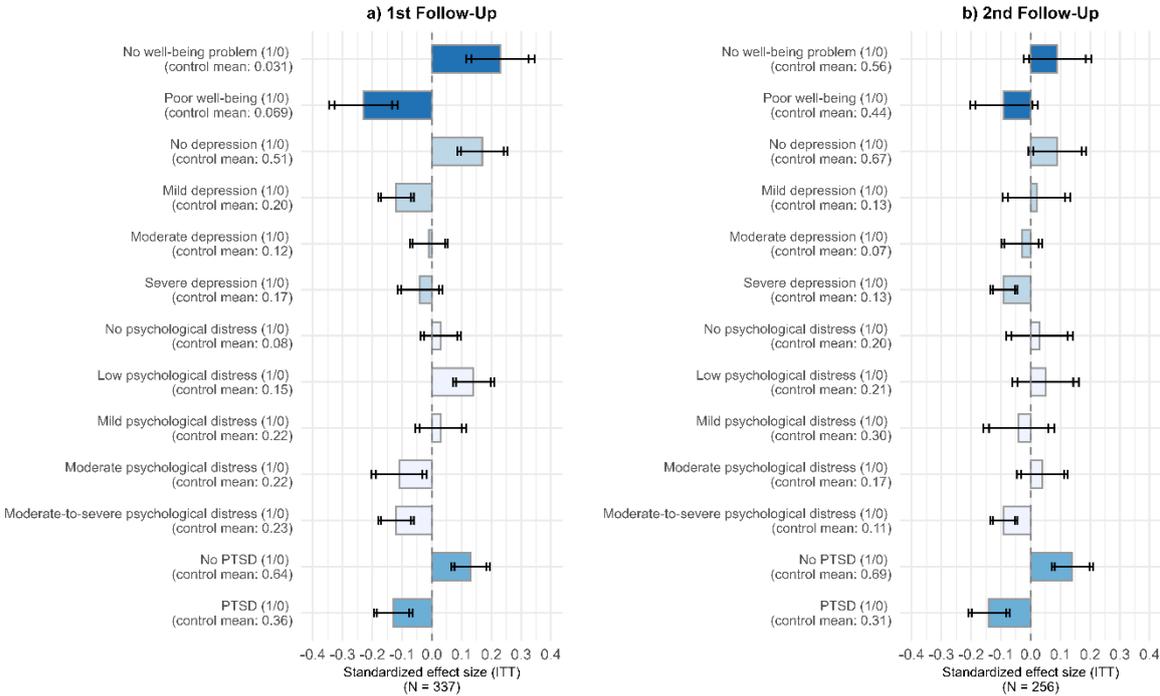

*Notes:* We show the impact of the intervention on respondents' level of psychological health. In Panel A, we show the short-term impact measured 12 months after treatment respondents were offered the intervention, and just before control respondents were given the option to begin. In Panel B, we report on the longer-term impact measured three to four years after treatment respondents were offered to start the intervention and two to three years after control respondents were offered to start theirs. Coefficients capture the effect on the share of respondents falling in category using our main specification. Confidence intervals are obtained using randomization inference with 2,000 replications. We have approximately 337 respondents in round 1 and 256 in round 2. See more information in Table A3.

# Tables

**Table 1:** Characteristics of the study respondents

| | (1) | (2) | (3) | (4) | (5) | (6) | (7) | (8) | (9) |
|---|---|---|---|---|---|---|---|---|---|
| | Baseline sample | | 1st Follow-up sample | | 2nd Follow-up sample | | (1)=(3) | (1)=(5) | (3)=(5) |
| | Mean | S.d. | Mean | S.d. | Mean | S.d. | P-val | P-val | P-val |
| *Panel A: Background characteristics* | | | | | | | | | |
| Male (1/0) | 0.26 | (0.44) | 0.26 | (0.44) | 0.27 | (0.44) | 0.90 | 0.82 | 0.92 |
| Age (year) | 30.01 | (3.09) | 30.07 | (3.08) | 29.87 | (3.04) | 0.83 | 0.56 | 0.43 |
| Years of education | 15.11 | (2.42) | 15.20 | (2.36) | 15.30 | (2.36) | 0.63 | 0.36 | 0.64 |
| In relationship (1/0) | 0.66 | (0.48) | 0.66 | (0.47) | 0.65 | (0.48) | 0.95 | 0.84 | 0.79 |
| Non-Danish ethnicity (1/0)) | 0.05 | (0.21) | 0.05 | (0.21) | 0.05 | (0.22) | 0.86 | 0.72 | 0.85 |
| Married (1/0) | 0.12 | (0.33) | 0.12 | (0.33) | 0.12 | (0.33) | 0.95 | 0.96 | 0.99 |
| Divorced (1/0) | 0.05 | (0.21) | 0.05 | (0.21) | 0.05 | (0.22) | 1.00 | 0.85 | 0.85 |
| Having children (1/0) | 0.24 | (0.43) | 0.24 | (0.43) | 0.23 | (0.42) | 0.94 | 0.79 | 0.85 |
| Number of children | 0.37 | (0.73) | 0.36 | (0.72) | 0.35 | (0.70) | 0.94 | 0.78 | 0.84 |
| *Panel B: Information on the problem for which help is sought* | | | | | | | | | |
| Age at which they became aware of the problem (year) | 10.98 | (5.12) | 11.04 | (5.09) | 11.19 | (5.17) | 0.89 | 0.64 | 0.74 |
| Adult(s) with a drug abuse problem (1/0) | 0.37 | (0.48) | 0.36 | (0.48) | 0.38 | (0.49) | 0.92 | 0.80 | 0.73 |
| Experienced violence during childhood (1/0) | 0.80 | (0.40) | 0.80 | (0.40) | 0.81 | (0.40) | 0.94 | 0.83 | 0.89 |
| Ever attempted suicide (1/0) | 0.14 | (0.35) | 0.15 | (0.35) | 0.14 | (0.35) | 0.91 | 0.95 | 0.87 |
| Family member attempted suicide (1/0) | 0.37 | (0.48) | 0.36 | (0.48) | 0.36 | (0.48) | 0.98 | 0.86 | 0.88 |
| *Panel C: Psychological health (dummies)* | | | | | | | | | |
| Poor mental well-being (1/0) | 0.62 | (0.49) | 0.62 | (0.49) | 0.61 | (0.49) | 1.00 | 0.79 | 0.79 |
| Mild to severe psychological distress (1/0) | 0.84 | (0.37) | 0.84 | (0.36) | 0.84 | (0.37) | 0.99 | 0.96 | 0.97 |
| Mild to severe depression (1/0) | 0.51 | (0.50) | 0.51 | (0.50) | 0.51 | (0.50) | 0.86 | 0.95 | 0.92 |
| PTSD (1/0) | 0.38 | (0.49) | 0.39 | (0.49) | 0.39 | (0.49) | 0.93 | 0.92 | 0.98 |
| *Panel D: Labor market status and benefits* | | | | | | | | | |
| Employment (1/0) | 0.64 | (0.48) | 0.65 | (0.48) | 0.65 | (0.48) | 0.84 | 0.84 | 1.00 |
| Full-time employment (1/0) | 0.16 | (0.37) | 0.16 | (0.37) | 0.15 | (0.35) | 0.95 | 0.62 | 0.67 |
| Avg. monthly earnings (USD) | 2,272.7 | (2,300.1) | 2,287.9 | (2,258.8) | 2,302.9 | (2,354.7) | 0.93 | 0.87 | 0.94 |
| Any benefits (1/0) | 0.50 | (0.50) | 0.50 | (0.50) | 0.51 | (0.50) | 0.91 | 0.87 | 0.79 |
| *- Unemployment benefits (1/0)* | 0.09 | (0.29) | 0.09 | (0.29) | 0.09 | (0.29) | 0.90 | 0.98 | 0.93 |
| *- Social assistance (1/0)* | 0.04 | (0.19) | 0.03 | (0.17) | 0.03 | (0.18) | 0.63 | 0.74 | 0.91 |
| *- Education benefits (1/0)* | 0.24 | (0.42) | 0.23 | (0.42) | 0.25 | (0.43) | 1.00 | 0.67 | 0.67 |
| *- Sickness benefits (1/0)* | 0.05 | (0.23) | 0.05 | (0.22) | 0.04 | (0.20) | 0.88 | 0.42 | 0.51 |
| *Panel E: Health* | | | | | | | | | |
| Avg. monthly usage of psycholeptics (DDD) | 0.64 | (2.87) | 0.63 | (2.92) | 0.80 | (3.27) | 0.97 | 0.52 | 0.50 |
| Avg. monthly usage of psychoanaleptics (DDD) | 4.00 | (13.30) | 3.97 | (13.40) | 3.75 | (12.85) | 0.98 | 0.81 | 0.84 |
| Usage of primary sector (count) | 19.83 | (14.41) | 19.94 | (14.53) | 19.28 | (14.10) | 0.92 | 0.64 | 0.58 |
| Visits to GP (count) | 12.78 | (9.89) | 12.77 | (9.99) | 12.42 | (9.79) | 0.98 | 0.66 | 0.68 |
| Visits to psychologist (count) | 1.04 | (3.08) | 1.00 | (2.99) | 1.04 | (3.04) | 0.85 | 0.98 | 0.89 |
| Visits to dermato-venerology clinic (count) | 0.43 | (1.49) | 0.44 | (1.52) | 0.33 | (1.21) | 0.92 | 0.37 | 0.33 |
| Has a problem w/ alcohol (CAGE-C) | 0.19 | (0.39) | 0.19 | (0.40) | 0.17 | (0.38) | 0.92 | 0.57 | 0.51 |

*Notes:* In this table, we provide the average characteristics of respondents included in our sample at baseline, of respondents who completed the first follow-up questionnaire, and of respondents who completed the second follow-up questionnaire. In columns (10), (11), and (12), we test the comparability of the characteristics of the individuals included in the three samples. Variables in panel A to D is measured in the month prior to randomization where the health information in Panel E are measured in the six-month leading up to randomization. The sample consists of approximately 358 respondents at baseline, 337 respondents in Round 1, and 256 respondents in Round 2.

**Table 2:** Balance checks

| | (1) | (2) | (3) | (4) | (5) | (6) | (7) | (8) | (9) |
|---|---|---|---|---|---|---|---|---|---|
| | Baseline sample | | | 1st Follow-up sample | | | 2nd Follow-up sample | | |
| | Control mean | Diff. | P-val | Control mean | Diff. | P-val | Control mean | Diff. | P-val |
| Survey data attrition (1/0) | | | | 0.06 | 0.01 | (0.84) | 0.32 | -0.08 | (0.10) |
| *Panel A: Background characteristics* | | | | | | | | | |
| Male (1/0) | 0.23 | 0.04 | (0.37) | 0.23 | 0.06 | (0.23) | 0.25 | 0.03 | (0.63) |
| Age (year) | 30.14 | -0.25 | (0.43) | 30.08 | -0.05 | (0.89) | 29.89 | -0.03 | (0.93) |
| Years of education | 15.10 | 0.04 | (0.89) | 15.17 | 0.06 | (0.80) | 15.14 | 0.36 | (0.22) |
| In relationship (1/0) | 0.65 | 0.00 | (1.00) | 0.65 | 0.02 | (0.82) | 0.63 | 0.04 | (0.60) |
| Non-Danish ethnicity (1/0)) | 0.04 | 0.01 | (0.69) | 0.04 | 0.01 | (0.78) | 0.05 | 0.02 | (0.57) |
| Married (1/0) | 0.10 | 0.04 | (0.32) | 0.10 | 0.04 | (0.27) | 0.12 | 0.01 | (0.78) |
| Divorced (1/0) | 0.06 | -0.02 | (0.43) | 0.06 | -0.02 | (0.31) | 0.06 | -0.01 | (0.61) |
| Having children (1/0) | 0.26 | -0.03 | (0.53) | 0.24 | -0.01 | (0.93) | 0.25 | -0.04 | (0.52) |
| Number of children | 0.38 | -0.04 | (0.59) | 0.36 | 0.00 | (0.99) | 0.38 | -0.06 | (0.52) |
| *Panel B: Information on the problem for which help is sought* | | | | | | | | | |
| Age at which they became aware of the problem (year) | 11.21 | -0.46 | (0.43) | 11.26 | -0.37 | (0.54) | 10.93 | 0.53 | (0.44) |
| Adult(s) with a drug abuse problem (1/0) | 0.41 | -0.09* | (0.07) | 0.41 | -0.09* | (0.09) | 0.46 | -0.16** | (0.01) |
| Experienced violence during childhood (1/0) | 0.78 | 0.04 | (0.35) | 0.78 | 0.04 | (0.32) | 0.78 | 0.04 | (0.38) |
| Ever attempted suicide (1/0) | 0.15 | -0.00 | (0.97) | 0.15 | -0.00 | (0.98) | 0.13 | 0.02 | (0.63) |
| Family member attempted suicide (1/0) | 0.39 | -0.04 | (0.40) | 0.38 | -0.03 | (0.55) | 0.39 | -0.07 | (0.23) |
| *Panel C: Psychological health (std.)* | | | | | | | | | |
| Well-being (WHO5) | -0.06 | 0.11 | (0.31) | -0.09 | 0.16 | (0.16) | -0.03 | 0.12 | (0.35) |
| Psychological distress (CORE) | 0.05 | -0.10 | (0.35) | 0.07 | -0.13 | (0.21) | 0.04 | -0.15 | (0.22) |
| Depression (MDI) | 0.03 | -0.06 | (0.59) | 0.05 | -0.10 | (0.34) | 0.04 | -0.10 | (0.42) |
| PTSD (ITQ) | 0.04 | -0.09 | (0.43) | 0.07 | -0.15 | (0.18) | 0.14 | -0.19 | (0.13) |
| *Panel D: Labor market status and benefits* | | | | | | | | | |
| Employment (1/0) | 0.65 | -0.02 | (0.71) | 0.66 | -0.02 | (0.69) | 0.67 | -0.03 | (0.58) |
| Full-time employment (1/0) | 0.16 | 0.00 | (1.00) | 0.16 | 0.00 | (0.99) | 0.17 | -0.04 | (0.34) |
| Avg. monthly earnings (USD) | 2,319.4 | -99.4 | (0.69) | 2,327.7 | -79.8 | (0.76) | 2,450.1 | -251.6 | (0.38) |
| Any benefits (1/0) | 0.49 | 0.03 | (0.52) | 0.48 | 0.03 | (0.62) | 0.47 | 0.08 | (0.21) |
| *Panel E: Health* | | | | | | | | | |
| Avg. monthly usage of psycholeptics (DDD) | 0.57 | 0.12 | (0.70) | 0.59 | 0.05 | (0.88) | 0,66 | 0.26 | (0.53) |
| Avg. monthly usage of psychoanaleptics (DDD) | 2.35 | 3.25** | (0.02) | 2.37 | 3.36** | (0.02) | 2.04 | 3.02* | (0.07) |
| Usage of primary sector (count) | 19.49 | 0.59 | (0.71) | 19.55 | 0.75 | (0.64) | 18.97 | 0.39 | (0.82) |
| Visits to GP (count) | 12.70 | 0.13 | (0.91) | 12.71 | 0.10 | (0.93) | 12.39 | 0.06 | (0.96) |
| Visits to psychologist (count) | 0.85 | 0.39 | (0.23) | 0.79 | 0.42 | (0.21) | 0.68 | 0.60 | (0.11) |
| Visits to dermato-venerology clinic (count) | 0.49 | -0.12 | (0.47) | 0.49 | -0.11 | (0.54) | 0.39 | -0.16 | (0.33) |
| Has a problem w/ alcohol (CAGE-C) | 0.18 | 0.02 | (0.60) | 0.18 | 0.03 | (0.50) | 0.15 | 0.04 | (0.50) |

*Notes:* The table presents the average characteristics of control respondents in the sample, as well as those who completed the first and second follow-up surveys. For each sample, we test for difference between the treatment and control groups. To do this, we estimate the equation (1) with only strata indicators for each variable listed in the left column. Randomization inference with 2,000 random permutations p-values are reported in parenthesis. Variables in Panel A to D are measured in the month prior to randomization where the health information in Panel E are measured in the six-month period leading up to randomization. The sample consists of approximately 358 respondents at baseline, 337 respondents in Round 1, and 256 respondents in Round 2.
*, **, *** denote significance at the 10, 5, and 1 percent levels respectively.

**Table 3:** Participation in treatment and behavior of individuals on waiting lists timeline

|  | (1) | (2) | (3) |
|---|---|---|---|
|  |  | **With covariates** |  |
|  | Control mean | ITT | P-value |
| *Panel A: At first follow-up survey* | | | |
| **1) Admin data (N=358)** | | | |
| Received help from TUBA (1/0) | 0.06 | 0.94*** | (0.00) |
| **2) Survey data (N=337)** | | | |
| Received any help (1/0) | 0.50 | 0.50*** | (0.00) |
| Received help from TUBA (1/0) | 0.05 | 0.95*** | (0.00) |
| Received help outside of TUBA (1/0) | 0.47 | -0.19*** | (0.00) |
| *- Formal help (1/0)* | 0.33 | -0.14*** | (0.00) |
| *- Informal help (1/0)* | 0.30 | -0.11** | (0.02) |
| *Panel B: At second follow-up survey* | | | |
| **1) Admin data (N=358)** | | | |
| Received help from TUBA (1/0) | 0.89 | 0.12*** | (0.00) |
| *- Number of sessions* | 19.09 | 0.73 | (0.53) |
| *- Duration of intervention (days)* | 544.6 | -34.4 | (0.28) |
| **2) Survey data (N=256)** | | | |
| Received any help (1/0) | 1.00 | 0.00 | (1.00) |
| Received help from TUBA (1/0) | 1.00 | 0.00 | (1.00) |
| Received help outside of TUBA (1/0) | 0.45 | -0.01 | (0.92) |
| *- Formal help (1/0)* | 0.26 | 0.08 | (0.18) |
| *- Informal help (1/0)* | 0.26 | -0.05 | (0.33) |

*Notes*: This table presents the impact of waiting time on participants' levels of psychological support. Panel A focuses on the period between randomization and the first follow-up survey, conducted 12 months after randomization, during which participants in the control group were placed on a waiting list. Panel B covers the period between randomization and the second follow-up survey, conducted three to four years after randomization, by which time all control group participants had been offered the opportunity to start the intervention. To do so, we use TUBA administrative records to measure intervention take-up, the number of TUBA sessions attended, and the duration (in days) between the first and last therapy session (N = 358). Help-seeking behavior is measured using survey data, with samples of N = 337 at the first follow-up and N = 256 at the second follow-up.

For each outcome displayed in the left column of the table, we estimate equation (1) and report the intention-to-treat (ITT) estimates, along with p-values obtained via randomization inference based on 2,000 permutations. Column 1 reports the average value of the outcomes in the control group.

*, **, *** denote significance at the 10, 5, and 1 percent levels respectively.

**Table 4:** Impact of the intervention on respondents' psychological health (indexes)

|  | (1) | (2) | (3) | (4) | (5) | (6) |
|---|---|---|---|---|---|---|
|  |  |  | \multicolumn{2}{c}{W/o covariates} | \multicolumn{2}{c}{With covariates} |
|  | Raw control mean (not used for regression) | Standardized control mean (used for regression) | ITT | P-value | ITT | P-value |
| *Panel A: First follow-up survey (N=337)* | | | | | | |
| Mental well-being (WHO-5 score) | 46.13 | 0.00 | 0.60*** | (0.00) | 0.60*** | (0.00) |
| Psychological distress (CORE score) | 15.79 | 0.00 | -0.45*** | (0.00) | -0.39*** | (0.00) |
| Depression (MDI score) | 20.56 | 0.00 | -0.46*** | (0.00) | -0.40*** | (0.00) |
| Post-traumatic stress disorder (ITQ score) | 10.69 | 0.00 | -0.40*** | (0.00) | -0.32*** | (0.00) |
| *Panel B: Second follow-up survey (N=256)* | | | | | | |
| Mental well-being (WHO-5 score) | 56.83 | 0.00 | 0.25* | (0.05) | 0.22* | (0.09) |
| Psychological distress (CORE score) | 12.39 | 0.00 | -0.21* | (0.09) | -0.15 | (0.15) |
| Depression (MDI score) | 16.64 | 0.00 | -0.30** | (0.01) | -0.28*** | (0.01) |
| Post-traumatic stress disorder (ITQ score) | 9.07 | 0.00 | -0.28** | (0.02) | -0.19* | (0.08) |

*Notes:* This table shows the impact of eliminating waiting time on psychological health. In Panel A, we report the short-term impact measured 12 months after the treatment participants were offered the intervention and just before the control participants were given the option to begin. In Panel B, we show the longer-term impact, measured three to four years after treatment participants were offered the intervention, and two to three years after the control participants were given the option to start theirs. For each outcome displayed in the left column of the table, we estimate equation (1) and report the intention-to-treat (ITT) estimates, along with p-values obtained via randomization inference based on 2,000 permutations. Column 1 reports the mean of the outcome variable in the control group prior to standardization for descriptive purposes. All regressions, however, are estimated using standardized outcome variables. The sample consists of 337 respondents in Round 1 and 256 respondents in Round 2.
*, **, *** denote significance at the 10, 5, and 1 percent levels respectively.

**Table 5:** Impact of the intervention on respondents' labor market and health outcomes

|  | (1) | (2) | (3) | (4) | (5) |
|---|---|---|---|---|---|
|  |  | W/o covariates | | With covariates | |
|  | Raw control mean (used for regression) | ITT | P-value | ITT | P-value |
| *Panel A: First follow-up survey (N=353)* | | | | | |
| **Labor** | | | | | |
| Any benefits (1/0) | 0.47 | -0.01 | (0.79) | -0.03 | (0.53) |
| Employment (1/0) | 0.63 | 0.03 | (0.56) | 0.04 | (0.30) |
| Full-time employment (1/0) | 0.32 | -0.01 | (0.92) | -0.01 | (0.73) |
| Earnings (USD) | 2396.60 | -53.80 | (0.82) | -40.40 | (0.81) |
| **Health** | | | | | |
| Usage of psycholeptics (DDD) | 3.07 | 1.60 | (0.45) | 0.48 | (0.79) |
| Usage of psychoanaleptics (DDD) | 27.11 | 11.09 | (0.29) | -5.48 | (0.55) |
| Visits to GP (count) | 7.01 | -0.16 | (0.82) | 0.05 | (0.94) |
| Visits to psychologist (count) | 0.44 | -0.09 | (0.59) | -0.24 | (0.15) |
| Has a problem w/ alcohol (CAGE-C) | 0.23 | -0.07* | (0.08) | -0.07* | (0.05) |
| *Panel B: Second follow-up survey (N=353)* | | | | | |
| **Labor** | | | | | |
| Any benefits (1/0) | 0.29 | -0.01 | (0.79) | -0.01 | (0.82) |
| Employment (1/0) | 0.74 | 0.01 | (0.88) | -0.00 | (0.94) |
| Full-time employment (1/0) | 0.42 | 0.02 | (0.74) | 0.00 | (0.97) |
| Earnings (USD) | 3309.68 | 234.74 | (0.41) | 194.38 | (0.42) |
| **Health** | | | | | |
| Usage of psycholeptics (DDD) | 5.33 | -0.97 | (0.73) | -0.98 | (0.73) |
| Usage of psychoanaleptics (DDD) | 33.98 | 11.90 | (0.35) | 0.69 | (0.96) |
| Visits to GP (count) | 6.49 | -0.04 | (0.95) | 0.19 | (0.77) |
| Visits to psychologist (count) | 0.31 | 0.23 | (0.26) | 0.19 | (0.34) |
| Has a problem w/ alcohol (CAGE-C) | 0.42 | 0.08 | (0.45) | 0.08 | (0.45) |

*Notes*: This table shows the impact of eliminating waiting time for psychological support on participants' labor market and health outcomes at the 1st follow-up survey (Panel A), as well as the impact of receiving the intervention earlier on with Panel B showing outcomes at the 2nd follow-up survey. For each outcome displayed in the left column of the table, we estimate equation (1) and report the intention-to-treat (ITT) estimates, along with p-values obtained via randomization inference based on 2,000 permutations. Column 1 reports the average value of the outcomes in the control group.

Labor outcomes are measured in the month of the follow-up survey whereas the health information are measured in the six months leading up to the follow-up survey. Outcomes are measured using Danish register data; the sample includes approximately 353 respondents per round.

*, **, *** denote significance at the 10, 5, and 1 percent levels respectively.

**Table 6:** Heterogeneous impact of the intervention at the first follow-up survey

| | | (1) | (2) | (3) | (4) | (5) | (6) | (7) |
|---|---|---|---|---|---|---|---|---|
| | | Control mean | | Treatment effects | | | | (3)=(5) |
| Group 1 vs. Group 2 | | Group 1 | Group 2 | Group 1 | P-value | Group 2 | P-value | p-value |
| *Panel A: Well-being (standardized)* | | Raw (standardized) mean | | | | | | |
| (a) | Females vs. Males | 46.81 (0.04) | 43.79 (-0.13) | 0.59*** | (0.00) | 0.65*** | (0.01) | 0.82 |
| (b) | Low vs. high baseline well-being | 39.75 (-0.35) | 54.32 (0.44) | 0.65*** | (0.00) | 0.54*** | (0.00) | 0.59 |
| | Low vs. high baseline PTSD symptoms | 49.46 (0.18) | 42.99 (-0.17) | 0.49*** | (0.00) | 0.70*** | (0.00) | 0.32 |
| (c) | Without vs. with children at baseline | 47.57 (0.08) | 41.5 (-0.25) | 0.50*** | (0.00) | 0.92*** | (0.00) | 0.10* |
| | Below 30 vs. 30+ at baseline | 48.00 (0.10) | 43.95 (-0.12) | 0.47*** | (0.00) | 0.76*** | (0.00) | 0.16 |
| *Panel B: PTSD symptoms (standardized)* | | | | | | | | |
| (a) | Females vs. Males | 10.89 (0.03) | 10.00 (-0.12) | -0.33*** | (0.00) | -0.27 | (0.13) | 0.80 |
| (b) | Low vs. high baseline well-being | 11.27 (0.10) | 9.95 (-0.13) | -0.39*** | (0.00) | -0.25* | (0.09) | 0.47 |
| | Low vs. high baseline PTSD symptoms | 8.45 (-0.39) | 12.80 (0.37) | -0.35*** | (0.01) | -0.30** | (0.02) | 0.79 |
| (c) | Without vs. with children at baseline | 10.52 (-0.03) | 11.25 (0.10) | -0.28*** | (0.01) | -0.46** | (0.02) | 0.40 |
| | Below 30 vs. 30+ at baseline | 10.64 (-0.01) | 10.76 (0.01) | -0.34*** | (0.00) | -0.34** | (0.02) | 0.99 |
| *Panel C: Has a problem with alcohol* | | Raw mean | | | | | | |
| (a) | Females vs. Males | 0.22 | 0.28 | -0.06 | (0.18) | -0.12 | (0.18) | 0.60 |
| (b) | Low vs. high baseline well-being | 0.24 | 0.22 | -0.09 | (0.13) | -0.06 | (0.25) | 0.74 |
| | Low vs. high baseline PTSD symptoms | 0.25 | 0.22 | -0.13** | (0.01) | -0.02 | (0.70) | 0.18 |
| (c) | Without vs. with children at baseline | 0.26 | 0.15 | -0.07 | (0.16) | -0.09 | (0.25) | 0.79 |
| | Below 30 vs. 30+ at baseline | 0.24 | 0.22 | -0.11** | (0.01) | -0.03 | (0.68) | 0.34 |
| *Panel D: Has a full-time employment* | | | | | | | | |
| (a) | Females vs. Males | 0.30 | 0.39 | -0.01 | (0.81) | 0.00 | (0.99) | 0.87 |
| (b) | Low vs. high baseline well-being | 0.32 | 0.33 | -0.06 | (0.30) | 0.05 | (0.48) | 0.23 |
| | Low vs. high baseline PTSD symptoms | 0.35 | 0.30 | -0.00 | (0.99) | -0.01 | (0.86) | 0.89 |
| (c) | Without vs. with children at baseline | 0.34 | 0.27 | -0.01 | (0.76) | -0.01 | (0.90) | 0.97 |
| | Below 30 vs. 30+ at baseline | 0.35 | 0.30 | -0.04 | (0.45) | 0.04 | (0.49) | 0.31 |

*Notes:* This table presents the short-term effects of eliminating waiting time across various subgroups. We report impacts on mental well-being (Panel A), PTSD symptoms (Panel B), the likelihood of being identified as having an alcohol problem (Panel C), and the likelihood of holding full-time employment (Panel D). For each outcome, we estimate equation (2) to assess heterogeneity by (a) sex, (b) baseline psychological health, and (c) intention to have children. For each outcome listed in the left column, we report the estimated effects for each subgroup, along with p-values obtained via randomization inference based on 2,000 permutations. Columns 1 and 2 report the average value of the outcome in the control group for respondents in group 1 and group 2, respectively. When regressions are estimated using standardized outcome variables, the mean of the standardized outcome is reported in parentheses.
*, **, *** denote significance at the 10, 5, and 1 percent levels respectively.

**Table 7:** Heterogeneous impact of the intervention at the second follow-up survey

|  |  | (1) | (2) | (3) | (4) | (5) | (6) | (7) |
|---|---|---|---|---|---|---|---|---|
|  |  | Control mean | | Treatment effects | | | | (3)=(5) |
| Group 1 vs. Group 2 | | Group 1 | Group 2 | Group 1 | P-value | Group 2 | P-value | p-value |
| *Panel A: Well-being (standardized)* | | Raw (standardized) mean | | | | | | |
| (a) | Females vs. Males | 57.48 (0.04) | 54.75 (-0.12) | 0.18 | (0.24) | 0.42 | (0.11) | 0.46 |
| (b) | Low vs. high baseline well-being | 55.70 (-0.07) | 58.18 (0.08) | 0.14 | (0.46) | 0.39** | (0.04) | 0.37 |
|  | Low vs. high baseline PTSD symptoms | 58.26 (0.08) | 55.71 (-0.07) | 0.22*** | (0.01) | 0.29 | (0.11) | 0.77 |
| (c) | Without vs. with children at baseline | 56.88 (0.00) | 56.67 (-0.01) | 0.14 | (0.36) | 0.61** | (0.03) | 0.14 |
|  | Below 30 vs. 30+ at baseline | 57.89 (0.06) | 55.37 (-0.09) | 0.10 | (0.57) | 0.48** | (0.02) | 0.18 |
| *Panel B: PTSD symptoms (standardized)* | | | | | | | | |
| (a) | Females vs. Males | 9.54 (0.08) | 7.55 (-0.24) | -0.19 | (0.13) | -0.21 | (0.29) | 0.92 |
| (b) | Low vs. high baseline well-being | 8.53 (-0.08) | 9.71 (0.10) | -0.13 | (0.33) | -0.28* | (0.08) | 0.48 |
|  | Low vs. high baseline PTSD symptoms | 8.02 (-0.17) | 9.88 (0.13) | -0.36** | (0.01) | -0.02 | (0.91) | 0.10 |
| (c) | Without vs. with children at baseline | 8.13 (-0.15) | 11.90 (0.45) | -0.13 | (0.28) | -0.40 | (0.15) | 0.36 |
|  | Below 30 vs. 30+ at baseline | 8.71 (-0.06) | 9.55 (0.08) | -0.12 | (0.41) | -0.31* | (0.07) | 0.38 |
| *Panel C: Has a problem with alcohol* | | Raw mean | | | | | | |
| (a) | Females vs. Males | 0.42 | 0.41 | 0.08 | (0.51) | 0.04 | (0.84) | 0.87 |
| (b) | Low vs. high baseline well-being | 0.52 | 0.31 | 0.05 | (0.76) | 0.08 | (0.55) | 0.88 |
|  | Low vs. high baseline PTSD symptoms | 0.45 | 0.40 | -0.02 | (0.88) | 0.16 | (0.26) | 0.39 |
| (c) | Without vs. with children at baseline | 0.43 | 0.42 | 0.03 | (0.66) | -0.06 | (0.57) | 0.48 |
|  | Below 30 vs. 30+ at baseline | 0.46 | 0.37 | -0.04 | (0.77) | 0.22 | (0.23) | 0.25 |
| *Panel D: Has a full-time employment* | | | | | | | | |
| (a) | Females vs. Males | 0.42 | 0.44 | -0.06 | (0.28) | 0.21** | (0.04) | 0.02** |
| (b) | Low vs. high baseline well-being | 0.40 | 0.46 | -0.03 | (0.64) | 0.04 | (0.62) | 0.47 |
|  | Low vs. high baseline PTSD symptoms | 0.44 | 0.41 | -0.00 | (0.97) | 0.01 | (0.92) | 0.92 |
| (c) | Without vs. with children at baseline | 0.47 | 0.27 | 0.05 | (0.70) | 0.16 | (0.49) | 0.69 |
|  | Below 30 vs. 30+ at baseline | 0.44 | 0.40 | 0.04 | (0.53) | -0.03 | (0.68) | 0.43 |

*Notes:* This table presents the longer-term effects of eliminating waiting time across various subgroups. We report impacts on mental well-being (Panel A), PTSD symptoms (Panel B), the likelihood of being identified as having an alcohol problem (Panel C), and the likelihood of holding full-time employment (Panel D). For each outcome, we estimate equation (2) to assess heterogeneity by (a) sex, (b) baseline psychological health, and (c) intention to have children. For each outcome listed in the left column, we report the estimated effects for each subgroup, along with p-values obtained via randomization inference based on 2,000 permutations. Columns 1 and 2 report the average value of the outcome in the control group for respondents in group 1 and group 2, respectively. When regressions are estimated using standardized outcome variables, the mean of the standardized outcome is reported in parentheses.
*, **, *** denote significance at the 10, 5, and 1 percent levels respectively.

# Appendix

**Appendix figure:**

**Figure A1:** Study timeline

**Appendix tables:**

**Table A1:** Comparison of trial participants and non-participants in 2018

**Table A2:** Characteristics of control respondents who sought help while on the waiting list

**Table A3:** Impact of the intervention on respondents' psychological health (categories)

**Table A4a:** Impact of the intervention on respondents' psychological health (indexes), focusing specifically on respondents who answered both the 1st and 2nd follow-up survey

**Table A4b:** Impact of the intervention on respondents' psychological health (categories), focusing specifically on respondents who answered both the 1st and 2nd follow-up survey

# Appendix Figures

**Figure A1:** Study timeline

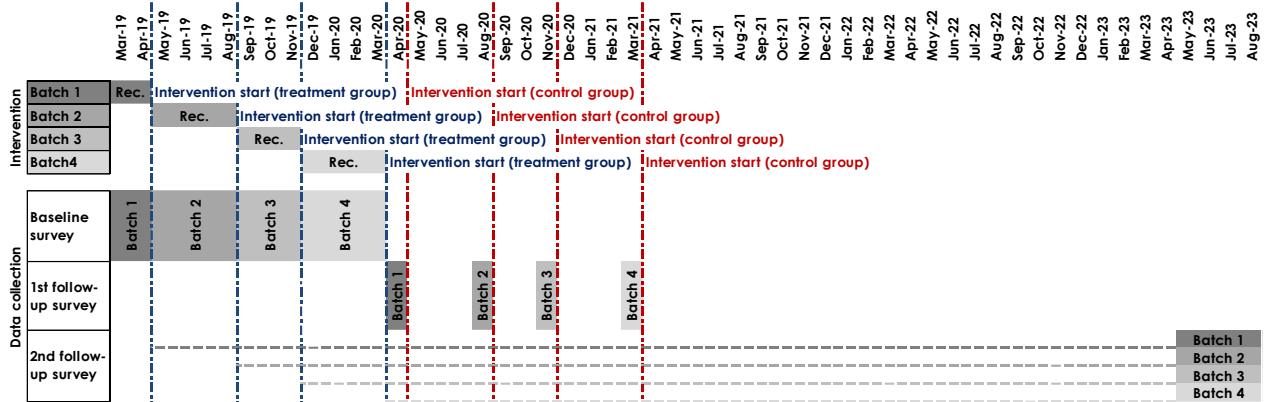

*Notes:* This figure shows the study timeline. Respondents were recruited in four batches over a 13-month period. In each batch, half of the respondents were assigned to a treatment group starting the intervention immediately, and the other half to a control group starting the intervention one year later. The first follow-up survey was conducted one year after randomization, before the start of the intervention for control group respondents (*i.e.* April 20 for batch 1 respondents, August 20 for batch 2 respondents, November 20 for batch 3 respondents and March 21 for batch 4 respondents). Between May and August 23, the second follow-up survey was carried out simultaneously for all four batches, 49 (37) months after intervention start for batch 1 treatment (control) respondents, 45 (33) months after intervention start for batch 2 treatment (control) respondents, 42 (30) months after intervention start for batch 3 treatment (control) respondents, and 38 (26) months after intervention start for batch 4 treatment (control) respondents.

# Appendix Tables

**Table A1:** Comparison of trial participants and non-participants in 2018

| | (1) | (2) | (3) | (4) | (5) | (6) | (7) | (8) |
|---|---|---|---|---|---|---|---|---|
| | Participants (study sample) | | Danish population (same age group) | | Difference | | Difference conditional on sex | |
| Variable | Mean | S.d. | Mean | S.d. | Coef. | S.e. | Coef. | S.e. |
| *Panel A: Background characteristics* | | | | | | | | |
| Female (1/0) | 0.74 | 0.44 | 0.49 | 0.50 | 0.25** | (0.02) | | |
| Age (years) | 28.83 | 3.16 | 29.63 | 4.30 | -0.80** | (0.17) | -0.81** | (0.17) |
| Non-Danish ethnicity (1/0) | 0.04 | 0.20 | 0.24 | 0.43 | -0.20** | (0.01) | -0.20** | (0.01) |
| Years of education | 14.32 | 2.87 | 13.51 | 4.14 | 0.81** | (0.15) | 0.68** | (0.15) |
| Married (1/0) | 0.11 | 0.32 | 0.27 | 0.44 | -0.15** | (0.02) | -0.17** | (0.02) |
| Have children (1/0) | 0.21 | 0.41 | 0.40 | 0.49 | -0.19** | (0.02) | -0.23** | (0.02) |
| Number of children | 0.32 | 0.69 | 0.69 | 0.98 | -0.37** | (0.04) | -0.44** | (0.04) |
| *Panel B: Labor market status* | | | | | | | | |
| Employment (months) | 5.66 | 4.86 | 6.94 | 4.92 | -1.27** | (0.26) | -1.02** | (0.26) |
| Annual earnings (USD) | 23,264.3 | 21,762.2 | 32,593.4 | 52,942.3 | -9,329.1** | (1,161.0) | -7,069.1** | (1,146.4) |
| Absence (days) | 8.79 | 32.47 | 4.35 | 19.36 | 4.44** | (1.73) | 3.45** | (1.74) |
| Not on public benefits (weeks) | 21.17 | 22.01 | 31.37 | 22.30 | -10.20** | (1.17) | -8.23** | (1.16) |
| *Panel C: Public benefits* | | | | | | | | |
| Unemployment benefits (weeks) | 4.26 | 10.63 | 2.15 | 7.41 | 2.11** | (0.57) | 1.95** | (0.57) |
| Social assistance (weeks) | 3.77 | 12.93 | 2.85 | 10.92 | 0.92 | (0.69) | 0.79 | (0.69) |
| Education benefits (weeks) | 17.82 | 22.64 | 8.94 | 18.18 | 8.88** | (1.21) | 8.26** | (1.20) |
| Sickness benefits (weeks) | 1.79 | 6.99 | 0.93 | 4.49 | 0.86** | (0.37) | 0.78** | (0.37) |
| *Panel D: Health care usage* | | | | | | | | |
| Primary health care usage (count) | 18.97 | 15.14 | 12.50 | 14.51 | 6.47** | (0.81) | 4.15** | (0.77) |
| Visits to dermatology and | 0.54 | 4.18 | 0.21 | 1.45 | 0.34 | (0.22) | 0.30 | (0.22) |
| Visits to dentist (count) | 2.14 | 2.56 | 1.78 | 2.56 | 0.36** | (0.14) | 0.28** | (0.14) |
| Visits to physiotherapy (count) | 1.01 | 3.80 | 0.98 | 5.29 | 0.03 | (0.20) | -0.07 | (0.20) |
| Visits to psychologist (count) | 0.82 | 3.10 | 0.25 | 1.67 | 0.56** | (0.17) | 0.52** | (0.17) |
| Visits to GP (count) | 13.06 | 11.85 | 8.24 | 10.20 | 4.82** | (0.63) | 2.97** | (0.60) |
| Primary health care costs (USD) | 378.63 | 421.12 | 235.76 | 365.94 | 142.88** | (22.45) | 101.15** | (21.92) |
| Number of observations | 351 | | 1,100,552 | | 1,100,903 | | | |

*Notes:* The table presents information on the average characteristics of respondents included in the sample (participants), as well as the characteristics of the Danish population within the same age group (non-participants)-individual born between 1981 and 1995, excluding those sampled. All variables are measured as of 2018. We test for differences between these groups by regressing each variable listed in the left column of the table on a dummy variable indicating whether an individual is part of the study sample. We also condition on sex to test if the differences are driven by different sex composition. Robust standard errors are computed.
*, **, and *** indicate significance at the 10, 5, and 1 percent levels respectively.

**Table A2:** Characteristics of control respondents who sought help while on the waiting list

|  | (1) Mean of respondents who did not receive formal help | (2) Difference between control respondents who received formal help and those who did not | (3) P-value |
|---|---|---|---|
| *Panel A: Background characteristics* | | | |
| Male (1/0) | 0.24 | -0.04 | (0.59) |
| Age (year) | 28.96 | 1.22 | (0.22) |
| Years of education | 14.70 | 0.39 | (0.51) |
| In relationship (1/0) | 0.64 | -0.04 | (0.66) |
| Married (1/0) | 0.11 | -0.06 | (0.24) |
| Having children (1/0) | 0.26 | -0.06 | (0.41) |
| Number of children | 0.39 | -0.08 | (0.49) |
| *Panel B: Information on the problem for which help is sought* | | | |
| Age at which they became aware of problem (year) | 9.91 | -0.01 | (1.00) |
| Adult(s) with a drug abuse problem (1/0) | 0.43 | -0.08 | (0.34) |
| Experienced violence during childhood (1/0) | 0.75 | 0.06 | (0.43) |
| Ever attempted suicide (1/0) | 0.13 | 0.05 | (0.39) |
| Family member attempted suicide (1/0) | 0.32 | 0.14* | (0.08) |
| *Panel C: Psychological health (std.)* | | | |
| Well-being (WHO5) | 0.07 | -0.48*** | (0.00) |
| Depression (MDI) | -0.11 | 0.49*** | (0.00) |
| Psychological distress (CORE) | -0.11 | 0.58*** | (0.00) |
| PTSD (ITQ) | -0.06 | 0.37** | (0.04) |
| *Panel D: Labor market status and benefits* | | | |
| Employment (1/0) | 0.70 | -0.21*** | (0.01) |
| Full-time employment (1/0) | 0.18 | -0.06 | (0.34) |
| Avg. monthly earnings (USD) | 2,559.3 | -844.6** | (0.04) |
| Any benefits (1/0) | 0.44 | 0.15* | (0.07) |
| *Panel F: Health* | | | |
| Avg. monthly usage of psycholeptics (DDD) | 0.53 | 0.10 | (0.82) |
| Avg. monthly usage of psychoanaleptics (DDD) | 2.14 | 0.72 | (0.65) |
| Usage of primary sector (count) | 17.39 | 4.92** | (0.05) |
| Visits to GP (count) | 11.99 | 2.00 | (0.20) |
| Visits to psychologist (count) | 0.30 | 1.29*** | (0.01) |
| Visits to dermato-venerology clinic (count) | 0.37 | 0.15 | (0.62) |
| Has a problem w/ alcohol (CAGE-C) | 0.18 | 0.00 | (1.00) |

*Notes:* In the table, we provide in column (2) the average characteristics of control respondents who did not receive any formal help. We also test for differences across formal vs no formal help. To do so, we estimate the difference between formal vs no formal help for each variable displayed in the left column. Randomization inference with 2,000 random permutations p-values are reported in parenthesis.
*, **, *** denote significance at the 10, 5, and 1 percent levels respectively.

**Table A3:** Impact of the intervention on respondents' psychological health (categories)

|  | (1) | (2) | (3) | (4) | (5) |
|---|---|---|---|---|---|
|  |  | W/o covariates | | With covariates | |
|  | Control mean | ITT | P-value | ITT | P-value |
| *Panel A: First follow-up survey* | | | | | |
| No mental well-being problem (1/0) | 0.31 | 0.23*** | (0.00) | 0.23*** | (0.00) |
| Poor mental well-being (1/0) | 0.69 | -0.23*** | (0.00) | -0.23*** | (0.00) |
| | | | | | |
| No psychological distress (1/0) | 0.08 | 0.05 | (0.11) | 0.03 | (0.38) |
| Low psychological distress (1/0) | 0.15 | 0.14*** | (0.00) | 0.14*** | (0.00) |
| Mild psychological distress (1/0) | 0.22 | 0.02 | (0.62) | 0.03 | (0.49) |
| Moderate psychological distress (1/0) | 0.27 | -0.10** | (0.02) | -0.11** | (0.02) |
| Moderate-to-severe psychological distress (1/0) | 0.23 | -0.13*** | (0.00) | -0.12*** | (0.00) |
| | | | | | |
| No depression (1/0) | 0.51 | 0.19*** | (0.00) | 0.17*** | (0.00) |
| Mild depression (1/0) | 0.20 | -0.12*** | (0.00) | -0.12*** | (0.00) |
| Moderate depression (1/0) | 0.12 | -0.01 | (0.80) | -0.01 | (0.76) |
| Severe depression (1/0) | 0.17 | -0.06 | (0.13) | -0.04 | (0.30) |
| | | | | | |
| No PTSD (1/0) | 0.64 | 0.15*** | (0.00) | 0.13*** | (0.00) |
| PTSD (1/0) | 0.36 | -0.15*** | (0.00) | -0.13*** | (0.00) |
| *Panel B: Second follow-up survey* | | | | | |
| No mental well-being problem (1/0) | 0.56 | 0.12* | (0.06) | 0.09 | (0.12) |
| Poor mental well-being (1/0) | 0.44 | -0.12* | (0.06) | -0.09 | (0.12) |
| | | | | | |
| No psychological distress (1/0) | 0.20 | 0.04 | (0.43) | 0.03 | (0.60) |
| Low psychological distress (1/0) | 0.21 | 0.05 | (0.35) | 0.05 | (0.38) |
| Mild psychological distress (1/0) | 0.30 | -0.05 | (0.41) | -0.04 | (0.51) |
| Moderate psychological distress (1/0) | 0.17 | 0.03 | (0.51) | 0.04 | (0.36) |
| Moderate-to-severe psychological distress (1/0) | 0.11 | -0.08*** | (0.01) | -0.09*** | (0.00) |
| | | | | | |
| No depression (1/0) | 0.67 | 0.10* | (0.07) | 0.09* | (0.07) |
| Mild depression (1/0) | 0.13 | -0.00 | (0.99) | 0.02 | (0.73) |
| Moderate depression (1/0) | 0.07 | -0.02 | (0.51) | -0.03 | (0.39) |
| Severe depression (1/0) | 0.13 | -0.08** | (0.01) | -0.09*** | (0.00) |
| | | | | | |
| No PTSD (1/0) | 0.69 | 0.17*** | (0.00) | 0.14*** | (0.00) |
| PTSD (1/0) | 0.31 | -0.17*** | (0.00) | -0.14*** | (0.00) |

*Notes:* This table shows the impact of eliminating waiting time for psychological support on participants' psychological health at 12 months (Panel A), as well as the impact of receiving the intervention earlier on (Panel B). For each outcome displayed in the left column of the table, we estimate equation (1) and report the intention-to-treat (ITT) estimates, along with p-values obtained via randomization inference based on 2,000 permutations. Column 1 reports the mean of the outcome variable in the control group. The sample consists of 337 respondents in Round 1, and 256 respondents in Round 2.
*, **, *** denote significance at the 10, 5, and 1 percent levels respectively.

**Table A4a:** Impact of the intervention on respondents' psychological health (indexes), focusing specifically on respondents who answered both the 1st and 2nd follow-up survey

|  | (1) | (2) | (3) | (4) | (5) |
|---|---|---|---|---|---|
|  |  | W/o covariates | | With covariates | |
|  | Standardized control mean | ITT | P-value | ITT | P-value |
| *Panel A: First follow-up survey* | | | | | |
| Mental well-being (WHO-5 score) | 0.06 | 0.59*** | (0.00) | 0.58*** | (0.00) |
| Psychological distress (CORE score) | -0.01 | -0.48*** | (0.00) | -0.42*** | (0.00) |
| Depression (MDI score) | -0.04 | -0.47*** | (0.00) | -0.38*** | (0.00) |
| Post-traumatic stress disorder (ITQ score) | 0.04 | -0.52*** | (0.00) | -0.38*** | (0.00) |
| *Panel B: Second follow-up survey* | | | | | |
| Mental well-being (WHO-5 score) | -0.02 | 0.26* | (0.05) | 0.23* | (0.06) |
| Psychological distress (CORE score) | 0.01 | -0.22* | (0.06) | -0.16 | (0.13) |
| Depression (MDI score) | 0.02 | -0.32*** | (0.01) | -0.29*** | (0.01) |
| Post-traumatic stress disorder (ITQ score) | 0.02 | -0.32*** | (0.01) | -0.22** | (0.04) |

*Notes*: In this table, we show the impact of eliminating waiting time for psychological support on participants' psychological health at 12 months (Panel A), as well as the impact of receiving the intervention earlier on (Panel B), specifically focusing on respondents answering both the 1st and 2nd follow-up survey. For each outcome displayed in the left column of the table, we estimate equation (1) and report the intention-to-treat (ITT) estimates, along with p-values obtained via randomization inference based on 2,000 permutations. Column 1 reports the standardized mean of the outcome variable in the control group. All regressions are estimated using standardized outcome variables. The sample consists of 251 respondents answering both surveys.

*, **, *** denote significance at the 10, 5, and 1 percent levels respectively.

**Table A4b:** Impact of the intervention on respondents' psychological health (categories), focusing specifically on respondents who answered both the 1st and 2nd follow-up survey

|  | (1) | (2) | (3) | (4) | (5) |
|---|---|---|---|---|---|
|  |  | W/o covariates | | With covariates | |
|  | Control mean | ITT | P-value | ITT | P-value |
| *Panel A: First follow-up survey* | | | | | |
| No mental well-being problem (1/0) | 0.32 | 0.26*** | (0.00) | 0.23*** | (0.00) |
| Poor mental well-being (1/0) | 0.68 | -0.26*** | (0.00) | -0.23*** | (0.00) |
|  |  |  |  |  |  |
| No psychological distress (1/0) | 0.09 | 0.06 | (0.25) | 0.02 | (0.68) |
| Low psychological distress (1/0) | 0.13 | 0.14*** | (0.01) | 0.17*** | (0.00) |
| Mild psychological distress (1/0) | 0.22 | 0.05 | (0.36) | 0.06 | (0.29) |
| Moderate psychological distress (1/0) | 0.29 | -0.14** | (0.01) | -0.13** | (0.02) |
| Moderate-to-severe psychological distress (1/0) | 0.23 | -0.12*** | (0.01) | -0.14*** | (0.00) |
|  |  |  |  |  |  |
| No depression (1/0) | 0.54 | 0.17*** | (0.01) | 0.15*** | (0.01) |
| Mild depression (1/0) | 0.17 | -0.07 | (0.14) | -0.07* | (0.10) |
| Moderate depression (1/0) | 0.14 | -0.05 | (0.32) | -0.04 | (0.33) |
| Severe depression (1/0) | 0.15 | -0.06 | (0.19) | -0.03 | (0.50) |
|  |  |  |  |  |  |
| No PTSD (1/0) | 0.62 | 0.21*** | (0.00) | 0.17*** | (0.00) |
| PTSD (1/0) | 0.38 | -0.21*** | (0.00) | -0.17*** | (0.00) |
| *Panel B: Second follow-up survey* | | | | | |
| No mental well-being problem (1/0) | 0.56 | 0.11* | (0.08) | 0.08 | (0.15) |
| Poor mental well-being (1/0) | 0.44 | -0.11* | (0.08) | -0.08 | (0.15) |
|  |  |  |  |  |  |
| No psychological distress (1/0) | 0.19 | 0.05 | (0.33) | 0.03 | (0.57) |
| Low psychological distress (1/0) | 0.21 | 0.04 | (0.44) | 0.04 | (0.39) |
| Mild psychological distress (1/0) | 0.29 | -0.04 | (0.49) | -0.03 | (0.56) |
| Moderate psychological distress (1/0) | 0.17 | 0.03 | (0.63) | 0.04 | (0.43) |
| Moderate-to-severe psychological distress (1/0) | 0.11 | -0.08** | (0.01) | -0.10*** | (0.00) |
|  |  |  |  |  |  |
| No depression (1/0) | 0.66 | 0.11* | (0.06) | 0.10* | (0.06) |
| Mild depression (1/0) | 0.13 | -0.01 | (1.00) | 0.01 | (0.87) |
| Moderate depression (1/0) | 0.07 | -0.02 | (0.60) | -0.03 | (0.40) |
| Severe depression (1/0) | 0.13 | -0.08** | (0.03) | -0.10*** | (0.01) |
|  |  |  |  |  |  |
| No PTSD (1/0) | 0.69 | 0.18*** | (0.00) | 0.15*** | (0.00) |
| PTSD (1/0) | 0.31 | -0.18*** | (0.00) | -0.15*** | (0.00) |

*Notes:* This table shows the impact of eliminating waiting time for psychological support on participants' psychological health at 12 months (Panel A), as well as the impact of receiving the intervention earlier on (Panel B), specifically focusing on respondents answering both the 1st and 2nd follow-up survey. For each outcome displayed in the left column of the table, we estimate equation (1) and report the intention-to-treat (ITT) estimates, along with p-values obtained via randomization inference based on 2,000 permutations. Column 1 reports the mean of the outcome variable in the control group. The sample consists of 251 respondents answering both surveys.
*, **, *** denote significance at the 10, 5, and 1 percent levels respectively.